\documentstyle[twoside,fleqn,espcrc2,psfig]{article}

\newcommand{\AmS}{{\protect\the\textfont2
  A\kern-.1667em\lower.5ex\hbox{M}\kern-.125emS}}

\newcommand{\agt}{ \mathop{}_{\textstyle \sim}^{\textstyle >} }
\newcommand{\alt}{ \mathop{}_{\textstyle \sim}^{\textstyle <} }
\newcommand{\text}[1]{{\rm \; #1}}
\newcommand{\epem}{e^+e^-}

\newcommand{\selectron}{\tilde{e}}
\newcommand{\smuon}{\tilde{\mu}}
\newcommand{\slepton}{\tilde{l}}

\newcommand{\LSP}{\tilde{\chi}^0_1}
\newcommand{\neutralinotwo}{\tilde{\chi}^0_2}

\newcommand{\susyU}{\widetilde{U}}

\title{New Probes of Supersymmetry Beyond the Minimal Framework
\thanks{Talk given at the 5th International Conference
on Supersymmetries in Physics (SUSY 97), Philadelphia, PA, 27-31 May
1997.}}
\author{
Jonathan L. Feng
\thanks{Research Fellow, Miller Institute for Basic Research in Science.}
\address{
Lawrence Berkeley National Laboratory and Department of Physics,\\
University of California, Berkeley, CA 94720, U.S.A.}
}

\begin{document}

\thispagestyle{empty}
\begin{abstract} 

\vspace*{-1.8in}
\noindent
{\normalsize
\begin{minipage}[t]{3in}
\begin{flushleft}
hep-ph/9708361 \\
August 1997
\end{flushleft}
\end{minipage}
\hfill
\begin{minipage}[t]{3in}
\begin{flushright}
LBNL--40404 \\
UCB--PTH--97/30 \\
\end{flushright}
\end{minipage}
}
\vspace*{1.5in}

If supersymmetry is discovered at future colliders, what can we learn?
While our appreciation of the variety of possible supersymmetric
models has grown tremendously in recent years, most attempts to answer
this question have been in the context of some simple and highly
restrictive framework, such as minimal supergravity.  In this talk I
describe new probes of phenomena that are generic in models beyond the
minimal framework.  These include tests of supersymmetric flavor and
CP violation and probes of kinematically inaccessible superparticle
sectors through ``super-oblique corrections.''  Such probes have wide
applicability to distinguishing models, from gravity- and
gauge-mediated theories to hybrid models and models with flavor
symmetries.  Examples of measurements at LEP II, the LHC, and the NLC
are given.
\end{abstract}

\maketitle

\section{INTRODUCTION}
\label{sec:introduction}

The discovery of supersymmetry (SUSY) in low energy experiments or in
current and future high energy colliders is at present a subject of
great interest.  It is important to bear in mind, however, that the
discovery of supersymmetry will be just the first step on the long
road toward determining the particular form of supersymmetry realized
in nature.  Such a program will require the precise measurement of
superparticle properties and will likely be the focus of experimental
high energy physics for decades.

While low energy experiments are certainly promising places to look
for the effects of new physics, they are unlikely to yield strong
bounds on SUSY parameters.  In fact, even the unambiguous
identification of supersymmetry as the source of new low energy
signals may be far from straightforward.  It is almost certain,
therefore, that input from high energy colliders will be required to
measure SUSY parameters and to thereby determine which of the many
possible SUSY models is viable.

\begin{figure}
\centerline{\psfig{file=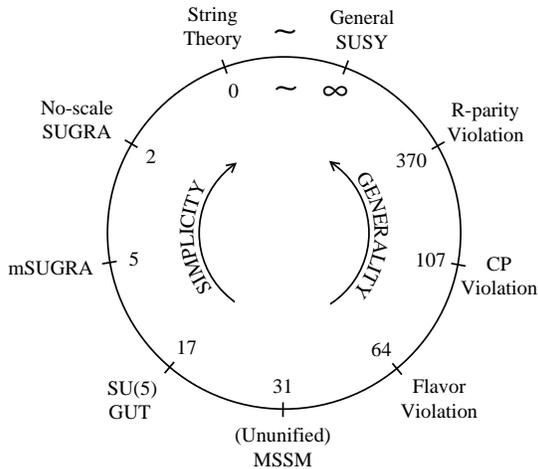,width=0.475\textwidth}}
\vspace*{-.3in}
\caption{Circle of Frameworks.  A sample of the possible theoretical 
frameworks for experimental studies and the number of accompanying new
SUSY parameters in each.}
\label{fig:circle}
\vspace*{-.2in}
\end{figure}

Given these considerations, an obvious question is, if SUSY is
discovered at colliders, what can we learn?  Despite the importance of
this question, the answer is, at the moment, far from thoroughly
understood.  The essential difficulty is that, although the low energy
supersymmetric theory is weakly coupled, and so observables are in
principle calculable to high precision, there are generically a large
number of parameters in supersymmetric models with even the most
minimal field content~\cite{DS}. For this reason, in any study of
possible experimental probes of SUSY, one must first choose a
theoretical framework. This is illustrated in Fig.~\ref{fig:circle},
where various theoretical SUSY frameworks are listed with the number
of new SUSY parameters accompanying them.
Minimal supergravity (mSUGRA) has its well-known 5 free parameters:
$m_0$, $m_{1/2}$, $A$, $\tan\beta$, and the sign of $\mu$.  However,
one may consider more general theories by successively relaxing
unification conditions and arriving at SU(5) grand unified theories,
with 17 free parameters, and the (ununified) MSSM, with 31.  If one
further relaxes flavor, CP, and R-parity conservation, the number of
parameters increases dramatically, and, of course, ultimately one can
consider models with non-minimal field content (General SUSY).  On the
other hand, one may consider more unified and highly constrained
frameworks until one arrives at string theory, which one might hope
would ultimately be a theory with no new parameters.  Unfortunately,
at present it is nearly equivalent to General SUSY in terms of its
predictive power for low energy SUSY parameters suitable for
experimental studies.

Fig.~\ref{fig:circle} is, of course, rather simplistic, and does not
include many other possible frameworks. The essential point, however,
is that there is a wide variety of theoretical frameworks, and any
choice of framework is necessarily a compromise between simplicity and
generality.  By far the most popular choice has been minimal
supergravity, the ``minimal framework'' referred to in the title of
this talk.  For example, the Snowmass '96 studies were conducted
almost exclusively in this framework~\cite{Bagger}.  While this is an
excellent and sometimes necessary starting point, it has led to a
lamentable, if understandable, dichotomy in the field of SUSY
phenomenology.  On the one hand, our understanding of the variety of
possible sparticle spectra, flavor structures, and SUSY breaking
mechanisms has grown tremendously in recent years, leading to a rapid
growth in the number of viable and appealing SUSY models.  At the same
time, much of the effort directed toward investigating the prospects
for studying SUSY at colliders has begun with the highly restrictive
assumption that the underlying theory is minimal supergravity.

In this talk, I will present a number of new probes of supersymmetric
models beyond the minimal framework. The motivations for relaxing the
strong assumptions of mSUGRA and considering less simple frameworks
are many, but a few of them may be listed here:

\noindent $\bullet$ 
Strong theoretical assumptions are inappropriate for studies hoping to
determine the prospects for testing theoretical assumptions.

\noindent $\bullet$  
Studies based on mSUGRA assumptions are fragile.  If a single
prediction of mSUGRA is disproven, studies based on mSUGRA must be
revised, or, worse, may become inapplicable.

\noindent $\bullet$ 
SUSY alone is already a highly constrained framework, and so strong
assumptions are not (all) necessary to obtain meaningful results.  For
example, studies have shown that superparticles with masses up to
$\sim 1$ TeV may be discovered at the LHC, even without the assumption
of R-parity conservation~\cite{BCT}.  Other studies find that gaugino
mass unification need not be assumed, but in fact may be tested both
at LEP II~\cite{Strassler} and future linear $\epem$
colliders~\cite{JLCI,FMPT}.

\noindent $\bullet$ 
Strong assumptions artificially suppress phenomena generic in SUSY.
For example, the scalar mass matrices are generically new sources of
flavor and CP violation in SUSY, but such violations are absent in the
minimal framework.  The presence (or the absence) of such phenomena
may in fact be powerful tools for excluding some models and favoring
others, as we will see below.

\noindent $\bullet$ 
Strong assumptions preclude important lessons for collider plans.  In
the present climate for building new colliders, it is essential that
general scenarios be considered in evaluating collider proposals.  As
we will see below, studies of SUSY beyond the minimal framework have
important lessons for collider parameters and options that are not
evident in the minimal framework.

At present, one of the leading motivations for models beyond minimal
supergravity is the desire to find new solutions to the supersymmetric
flavor problem, {\em i.e.}, the problem that low energy constraints
are violated for generic sfermion masses and mixings.  The problem may
be solved by sfermion degeneracy, fermion-sfermion alignment, very
massive sfermions, or some combination of these three.  In this talk,
I will describe a number of collider probes that may help determine
which solution is realized in nature, and may therefore be used to
eliminate some models in favor of others.  In Sec.~\ref{sec:mixing},
probes of scalar flavor structures at both $\epem$ and hadron machines
will be described.  In Sec.~\ref{sec:heavy}, I will discuss probes of
models with very heavy sparticles.  Such sparticles will not be
directly discovered at colliders, but their properties may be probed
through ``super-oblique parameters,'' supersymmetric analogues of the
oblique parameters of the standard model which will be introduced
below. This talk is an overview of work presented in
Refs.~\cite{lfv,cp,CFP1,CFP2}.

\section{SFERMION FLAVOR MIXING}
\label{sec:mixing}

If supersymmetric particles are discovered, measurements of their
masses will be of primary importance.  Such measurements have been
well-studied.  Their intergenerational mixings will also be of great
interest, however, especially given the importance of the SUSY flavor
problem.  In the standard model, all flavor mixing is confined to the
CKM matrix.  However, in any supersymmetric extension of the standard
model, both lepton and quark flavor are typically violated by
supersymmetric interactions. These new violations arise because the
scalar partners of the fermions must be given mass, and the scalar
mass matrices are generally not diagonal in the same basis as the
fermion masses.  There are then 7 new independent flavor
matrices\footnote{We neglect possible neutrino masses and treat
left-right scalar mixings as perturbations.}, $W_a$, $a=u_{L,R},
d_{L,R}, e_{L,R}, \nu_L$, which are analogues of the CKM matrix and
provide a rich variety of new phenomena that may be studied at
colliders.

Here we will focus on the leptonic sector. In the standard model,
lepton flavor is conserved, and so any lepton flavor violation
observed in the processes discussed below is necessarily
supersymmetric in origin.  The $W$ matrices appear in gaugino/Higgsino
vertices. For neutralinos $\tilde{\chi}^0$, these vertices are given
by the interactions
\begin{eqnarray}
\label{convention}
&&\tilde{e}_{Li} {W_L^*}_{i\alpha} \overline{e_{L\alpha}} 
\tilde{\chi}^0
+\tilde{e}^*_{Li} {W_L}_{i\alpha} \overline{{\tilde{\chi}}^0} 
e_{L\alpha}  \nonumber \\
&&+\tilde{e}_{Ri} {W_R^*}_{i\alpha} \overline{e_{R\alpha}} 
\tilde{\chi}^0 
+\tilde{e}^*_{Ri} {W_R}_{i\alpha} \overline{{\tilde{\chi}}^0} 
e_{R\alpha} \, ,
\end{eqnarray}
where the Latin and Greek subscripts are generational indices for
scalars and fermions, respectively.  For full three generation mixing,
the two matrices $W_L$ and $W_R$ may be parametrized by 6 mixing
angles and 4 phases.

\subsection{Flavor Violation}
\label{sec:flavor}

\begin{figure}
\centerline{\psfig{file=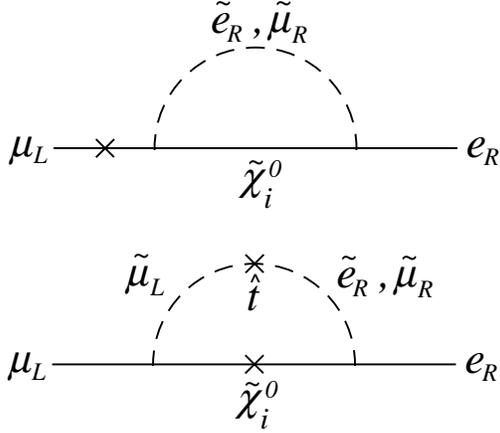,width=0.475\textwidth}}
\vspace*{-.5in}
\caption{The leading contributions to $\mu\to e\gamma$ in the gaugino
region.  An external photon is to be included in each diagram.}
\label{fig:mutoegamma}
\vspace*{-.2in}
\end{figure}

In this section we investigate probes of flavor mixing, and for
simplicity, we will set the CP-violating phases to zero and consider
two generation mixing.  It is important to remember that low energy
bounds already provide significant constraints on the $W$ matrices.
The most stringent constraints are from $B(\mu \to e \gamma) < 4.9
\times 10^{-11}$~\cite{LAMPF}, $B(\tau \to e \gamma) < 2.7 \times
10^{-6}$~\cite{CLEO}, and $B(\tau \to \mu \gamma) < 3.0 \times
10^{-6}$~\cite{CLEO}.  We will be ambitious by considering
$\selectron_R$--$\smuon_R$ mixing, which competes directly with the
strongest of these, the bound on $\mu \to e \gamma$. (13 and 23
mixing, as well as mixings among the left-handed sleptons, may also be
considered in analyses similar to the one described below.) The mixing
matrix $W_R$ may then be parametrized by a single mixing angle
$\theta_{12}$.  With this mixing, the leading contributions to $\mu
\to e \gamma$~\cite{BHS} in the gaugino region are from the 
two diagrams of Fig.~\ref{fig:mutoegamma}.  (In the mixed or Higgsino
regions, additional contributions reduce $B(\mu\to
e\gamma)$~\cite{HMTY}.)  Note that the rate depends on many SUSY
parameters, and so $\mu \to e \gamma$ alone will not provide
model-independent constraints on the SUSY parameters.  Both amplitudes
are proportional to $\sin 2\theta_{12}$ and are superGIM suppressed in
the limit of degenerate sleptons.  The second amplitude is also
proportional to the left-right mixing parameter $\hat{t}
\equiv (- A + \mu \tan\beta) / \bar{m}_{12}$.  The total rate then has
the form
\begin{equation}
B(\mu\to e\gamma) \sim f(\hat{t}\: \!) \left( \frac{\Delta
m_{12}^2}{\bar{m}_{12}^2}\sin 2\theta_{12} \right)^2 \, ,
\end{equation}
where $\bar{m}_{ij}^2 = (m_i^2 + m_j^2)/2$ is the average squared
slepton mass, and $\Delta m_{ij}^2 = (m_i^2 - m_j^2)/2 \approx 2m
\Delta m_{ij}$, with $\Delta m_{ij} = m_i-m_j$.  Note that the
superGIM mechanism suppresses the rate for $\Delta m < m$.

\begin{figure}
\centerline{\psfig{file=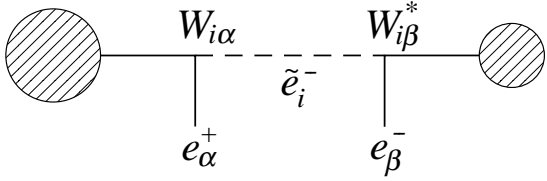,width=0.475\textwidth}}
\vspace*{-.3in}
\caption{A general lepton flavor-violating process involving 
single slepton production.}
\vspace*{-.2in}
\label{fig:mab}
\end{figure}

We now consider the rate for flavor-violating signals at colliders.
These signals arise from processes involving on-shell slepton
production with unlike flavor leptons in the final state. For
simplicity, let us consider processes involving a single
slepton.\footnote{The formalism for correlated slepton pair production
is more complicated and is presented in Ref.~\cite{cp}; however, the
essential conclusions given below remain unchanged.}  The general form
for such a process is $f_1 f_2 \to e^+_{\alpha} X
\selectron^-_i \to e^+_{\alpha} X e^-_{\beta} Y$, where $f_1$ and
$f_2$ are the initial state partons, and $X$ and $Y$ denote sets of
particles.  This process is depicted in Fig.~\ref{fig:mab}.
If we denote the amplitude of this reaction as ${\cal
M}^i_{\alpha\beta}$, the cross section receives contributions from
${\cal M}^i_{\alpha\beta}{\cal M}^{j \dagger}_{\alpha\beta}$, where
$i$ and $j$ are the generational indices of any sleptons that may be
produced in such a process.  The form for the flavor-violating cross
section in the presence of an arbitrary mixing matrix $W$ is
\begin{eqnarray}
\sigma_{\alpha\beta} &\equiv& \sigma
(f_1 f_2 \to e^+_{\alpha} X e^-_{\beta} Y ) \\
&=& \sigma_0
\sum_{ij} W_{i\alpha} W^*_{i\beta} W_{j\alpha}^* W_{j\beta} 
\frac{1}{1+ix_{ij}} \, , 
\label{sigmasingle} 
\end{eqnarray}
where $\sigma_0 = \sigma (f_1 f_2 \to e^+ X \selectron^-)
B(\selectron^- \to e^- Y)$ is the analogous cross section in the
absence of flavor violation, and, in analogy with the variables $x_d$
and $x_s$ in $B$ physics, we have defined $x_{ij}\equiv \Delta
m_{ij}/\Gamma$, where $\Gamma$ is the slepton decay width.  If the
mass splittings are much larger than the widths, $x_{ij}$ is large for
$i \ne j$ and only terms with $i=j$ contribute.  However, for $x_{ij}
\alt 1$, interference terms play an important role.

Let us now return to the simple case of two generation mixing.  In
this case, Eq.~(\ref{sigmasingle}) reduces to
\begin{equation}
\sigma_{12} = \sigma_0 \frac{x_{12}^2}{2(1+x_{12}^2)} 
\sin^2 2\theta_{12} \, .
\end{equation}
As with the low energy signal, the flavor-violating collider signal
vanishes in the limit of degenerate sleptons, as it must.  However, in
marked contrast with the low energy signal, the collider signal is
suppressed only for $\Delta m < \Gamma$.  There is thus a large region
of parameter space with $\Gamma \alt \Delta m \alt m$ in which the
low energy signal is highly suppressed, but the collider signal may be
large.
 
We now determine the reach of various colliders in the flavor mixing
parameter space.  We begin with LEP II, and assume an integrated
luminosity of 500 pb$^{-1}$ at $\sqrt{s} = 190 \text{ GeV}$.  To
present quantitative results, we choose slepton masses just beyond the
current bounds $\bar{m}_{12} \approx 80 \text{ GeV}$ and assume the
LSP is Bino-like with mass $M_1 = 50 \text{ GeV}$.  The effects of
variations from this scenario are discussed in Ref.~\cite{lfv}.

\begin{figure}
\centerline{\psfig{file=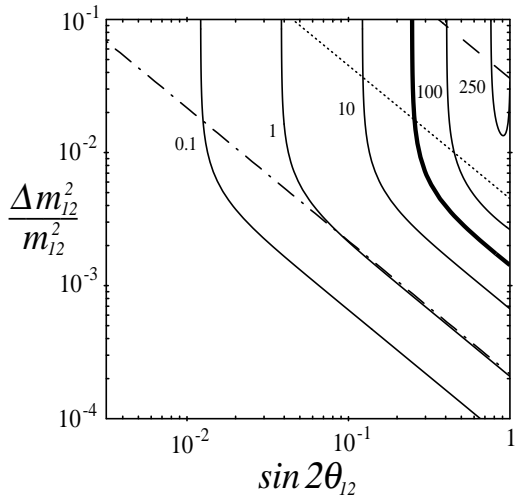,width=0.475\textwidth}}
\vspace*{-.3in}
\caption{Flavor Violation at LEP II.  Contours of constant $\sigma (
e^+e^- \to e^{\pm} \mu^{\mp} \tilde{\chi}^0\tilde{\chi}^0 )$ (solid)
in fb for $m_{\tilde{e}_R}, m_{\tilde{\mu}_R} \approx 80 \text{ GeV}$
and $M_1 = 50 \text{ GeV}$. The thick contour represents the optimal
experimental reach in one year (40 fb).  Constant contours of $B(\mu
\to e\gamma)=4.9\times 10^{-11}$ are also plotted for $\hat{t} \equiv
(- A + \mu \tan\beta)/\bar{m}_R = 0$ (dotted), 2 (dashed), and 50
(dot-dashed), and degenerate left-handed sleptons with $m_{\slepton_L}
\approx 120$ GeV.}
\label{fig:LEPII}
\vspace*{-.2in}
\end{figure}

With these choices, the flavor-violating signal we are looking for is
$\epem \to e^{\pm} \mu^{\mp} \LSP \LSP$.  The largest background is
$W$ pair production, but this may be reduced significantly with
appropriate cuts.  In Fig.~\ref{fig:LEPII}, flavor-violating cross
sections are given by the solid contours, with the 5$\sigma$ discovery
signal given by the thick contour.  Bounds from $\mu\to e\gamma$ are
also plotted --- note that the two diagrams of
Fig.~\ref{fig:mutoegamma} interfere destructively, and so $B(\mu\to
e\gamma)$ is not monotonic in $\hat{t}$ and even vanishes for certain
$\hat{t}$. We see that for low values of $\hat{t}$, LEP II extends the
reach in parameter space, and for mixing angles $\theta_{12}\agt
\theta_C$, where $\theta_C$ is the Cabibbo angle, lepton flavor
violation may be discovered at LEP II.  It is remarkable that for a
significant region in parameter space, LEP II, which will produce at
most a few hundred sleptons, is more sensitive to lepton flavor
violation than $\mu\to e\gamma$, with its astounding
statistics.\footnote{Of course, in the best case scenario in which
$\mu\to e \gamma$ is also discovered, both low and high energy
measurements will be useful for determining the flavor-violating
parameters.}  This example indicates the extremely promising prospects
for precision SUSY measurements once superpartners are discovered.

\begin{figure}
\centerline{\psfig{file=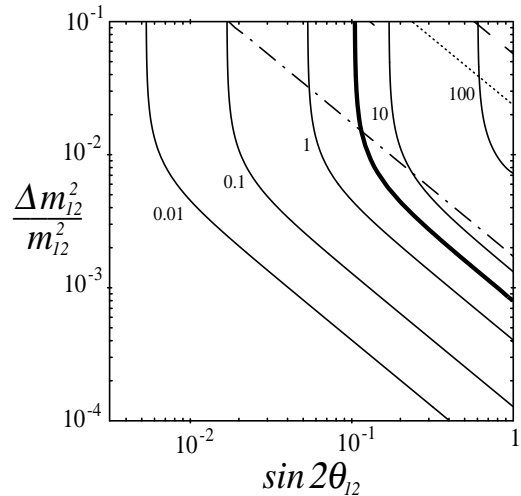,width=0.475\textwidth}}
\vspace*{-.3in}
\caption{Flavor Violation at the NLC.  Contours of constant 
$\sigma (e^+e^-_R\to e^{\pm} \mu^{\mp} \tilde{\chi}^0\tilde{\chi}^0 )$
(solid) in fb for $\protect\sqrt{s} = 500 \text{ GeV}$,
$m_{\tilde{e}_R}, m_{\tilde{\mu}_R} \approx 200 \text{ GeV}$, and $M_1
= 100 \text{ GeV}$ (solid).  The thick contour represents the
experimental reach in one year.  Constant contours of $B(\mu \to
e\gamma)$ are also plotted as in Fig.~\protect\ref{fig:LEPII}, but for
$m_{\slepton_L} \approx 350$ GeV.}
\label{fig:NLC+}
\vspace*{-.2in}
\end{figure}

At future linear colliders, the prospects for lepton flavor violation
discovery are even brighter.  Highly polarized $e^-$ beams may also be
used to reduce backgrounds.  In Fig.~\ref{fig:NLC+}, the discovery
reach at the NLC is shown for $\int {\cal L} = 50 \text{ fb}^{-1}$,
$\sqrt{s}= 500 \text{ GeV}$, a 90\% right-polarized $e^-$ beam, and
suitably scaled up SUSY parameters.
The NLC is sensitive to mixing angles $\theta_{12} \agt 0.05$, and
extends the reach in parameter space beyond current $\mu \to e \gamma$
bounds even for $\hat{t} = 50$. In fact, these bounds can be improved
still further by using the $e^-e^-$ mode of the NLC~\cite{lfv}.

\subsection{CP Violation}
\label{sec:CP}

In the presence of CP violation in the slepton mass matrix, the cross
sections $\sigma_{\alpha\beta} = \sigma (f_1 f_2 \to e^+_{\alpha} X
e^-_{\beta} Y)$ and $\sigma_{\beta\alpha} = \sigma (f_1 f_2 \to
e^+_{\beta} X e^-_{\alpha} Y)$ are no longer
equal~\cite{cp,BK}.\footnote{Phases in SUSY parameters such as $\mu$
and the gaugino masses may change these cross sections, but do not
distinguish between flavors, and so do not contribute to the
CP-violating difference.} It is easy to show from
Eq.~(\ref{sigmasingle}) that
\begin{eqnarray}
\lefteqn{\Delta_{\alpha\beta} \equiv 
\sigma_{\alpha\beta}-\sigma_{\beta\alpha}=-4\sigma_0 \times} 
\nonumber \\
\lefteqn{\quad \sum_{i<j}
Im \left[ W_{i\alpha} W^*_{i\beta} W_{j\alpha}^* W_{j\beta} \right] 
Im \left[ \frac{1}{1 + ix_{ij}} \right] \, .}
\label{CP}
\end{eqnarray}
Again, the sum is over sleptons that may be produced on-shell.  We
therefore reproduce the familiar result that CP violation requires the
presence of at least two amplitudes that differ both in their CP-odd
(``weak'') phases and their CP-even (``strong'') phases.  For what
parameters can this asymmetry be large? The slepton mass dependent
(CP-even) part is $-x_{ij}/(1+x_{ij}^2)$.  The $W$-dependent (CP-odd)
part may be written as
\begin{equation}
Im \left[ W_{i\alpha} W^*_{i\beta} W_{j\alpha}^* W_{j\beta} \right] 
= \widetilde{J} \sum_{k\gamma} \varepsilon_{ijk}
\varepsilon_{\alpha\beta\gamma} \, , 
\label{J}
\end{equation}
where $\widetilde{J}$ is the supersymmetric analogue to the Jarlskog
invariant, and is the unique re-phase invariant that may be formed
from a single $W$ matrix.  In a standard parametrization~\cite{Chau},
$\widetilde{J} = \sin\theta_{12} \cos\theta_{12} \sin\theta_{13}
\cos\theta_{13} \sin\theta_{23} \cos^2\theta_{23}\sin\delta$, 
where $\theta_{ij}$ are the 3 mixing angles, and $\delta$ is the
CP-violating phase.  We see then that the CP-violating collider signal
is maximal for $\Delta m \sim \Gamma$ and requires full 3 generation
mixing and, of course, a significant CP-violating phase $\delta$.

The two rephase invariants $\widetilde{J}_L$ and $\widetilde{J}_R$,
each formed from a single $W$ matrix, govern the sizes of CP-violating
collider signals. As in the case of flavor violation, it is important
to understand to what extent these signals are already bounded by low
energy constraints, such as, in this case, the electric dipole moment
(EDM) of the electron.  Recall, however, that there are 4 irremovable
phases in the two $W$ matrices. The EDM of the electron constrains the
rephase invariants $\widetilde{K}_{12} =
Im\left[{W_L}_{21}{W^*_L}_{22} {W^*_R}_{21}{W_R}_{22}\right]$ and
$\widetilde{K}_{13} = Im\left[{W_L}_{31}{W^*_L}_{33}
{W^*_R}_{31}{W_R}_{33}\right]$.  The four independent invariants
$\widetilde{J}_{L,R}$ and $\widetilde{K}_{12,13}$ therefore form a
convenient basis --- colliders probe $\widetilde{J}_{L,R}$ and the
electron EDM probes $\widetilde{K}_{12,13}$. Formally, we see that
electron EDM constraints provide no bounds on collider signals.  Of
course, it would be unnatural to expect a large hierarchy between the
$\widetilde{J}$'s and $\widetilde{K}$'s. However, one may check that
for $\Delta m \sim \Gamma$, where the collider probe may be relevant,
the bounds on the $\widetilde{K}$'s from $d_e < 4\times 10^{-27} e$
cm~\cite{edmref} are extremely weak.

We now consider possible probes of the CP-violating cross section
asymmetries at future colliders.  First, note that we are looking for
a statistically significant asymmetry $\Delta_{\alpha\beta}$ in a
sample of dilepton events. Contributions to this sample from
non-slepton events dilute this asymmetry, and so we must isolate a
large pure sample of slepton events.  Second, it is interesting that
Eqs.~(\ref{CP}) and (\ref{J}) imply that all three possible
asymmetries are equal, up to a sign: $|\Delta_{12}| = |\Delta_{13}| =
|\Delta_{23}|$.  Of course, backgrounds and other experimental issues
entering the measurements of these asymmetries differ from channel to
channel.  We will only consider the 12 asymmetry here, but note that
ultimately all three asymmetries may be combined to provide the most
powerful probe of $\widetilde{J}$.  Finally, as we must now consider
general three generation mixing, the parameter space is fairly
complicated.  To present our results, we characterize the mixing by a
single angle $\theta = \theta_{12} = \theta_{13} = \theta_{23}$, and
similarly, we fix $\Delta m = \Delta m_{12} = \Delta m_{23}$.
Variations from these assumptions may be found in Ref.~\cite{cp}.

\begin{figure}
\centerline{\psfig{file=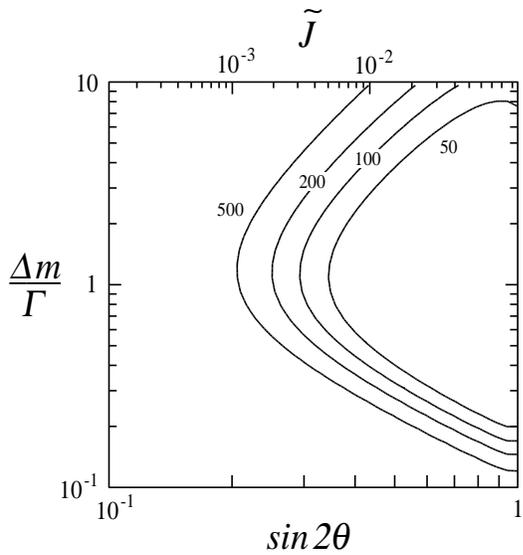,width=0.48\textwidth}}
\vspace*{-.3in}
\caption{CP Violation at the NLC.  3$\sigma$ slepton CP violation
discovery contours for the integrated luminosity given (in fb$^{-1}$).
The CP-violating phase is fixed to $\sin\delta = 1$.  The SUSY
parameters are as given in the text, with $\Gamma = 0.58\text{ GeV} =
2.9 \times 10^{-3} m$.
\label{fig:NLCI}}
\vspace*{-.2in}
\end{figure}

Let us begin by considering the NLC.  Again, we must choose parameters
to present quantitative results, and we consider the NLC parameters
chosen above with $m_{\slepton_R} \approx 200\text{ GeV}$.
Right-handed beam polarization and appropriate cuts effectively reduce
the number of dilepton events from the main backgrounds $WW$, $e\nu
W$, and $eeWW$.  The resulting 3$\sigma$ CP violation discovery
contours are presented in Fig.~\ref{fig:NLCI} for various integrated
luminosities.  We see that, for $\Delta m \approx \Gamma$,
$\widetilde{J}$ as low as $10^{-3}$ to $10^{-2}$ may be probed.  The
NLC provides a robust probe of slepton CP violation, for the most part
requiring only that slepton pairs be kinematically accessible.  In
addition, left- and right-handed slepton flavor and CP violation may
be disentangled by gradually raising the beam energy or using beam
polarization.

\begin{figure}[t]
\centerline{\psfig{file=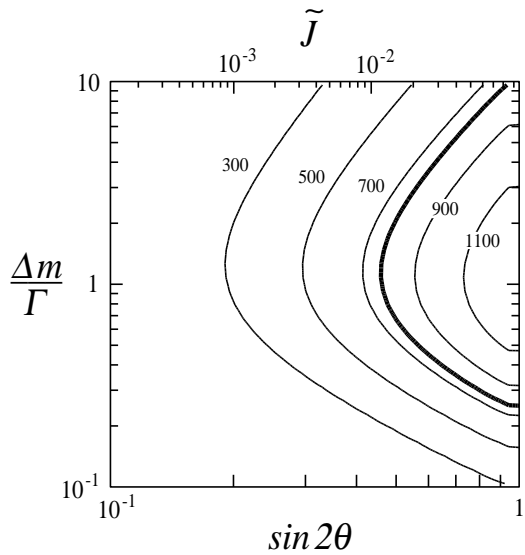,width=0.48\textwidth}}
\vspace*{-.3in}
\caption{CP Violation at the LHC. 3$\sigma$ slepton CP violation 
discovery contours for an integrated luminosity of 100 fb$^{-1}$.  The
CP-violating phase is fixed to $\sin\delta = 1$.  The wide contour is
for the parameters discussed in the text, where $m_{\tilde{g}}= 767$
GeV, and $\Gamma = 0.13\text{ GeV} = 8.1 \times 10^{-4} m$. The other
contours give an indication of possible discovery reaches for
scenarios with $m_{\tilde{g}} = m_{\tilde{q}}$ shown (in GeV).  See
full discussion in the text.
\label{fig:LHCI}}
\vspace*{-.25in}
\end{figure}

Probes of slepton CP violation are also possible at hadron machines if
a large sample of slepton events may be isolated.  At the LHC with
$\sqrt{s} = 14$ TeV, the most promising source of sleptons is in
cascade decays of squarks and gluinos.  To study this possibility, we
consider a scenario studied in Ref.~\cite{SnowmassLHC}, in which the
following cascade decays occur (sparticle masses in GeV and branching
ratios in each step are indicated): $\left[ \tilde{g} \ (767)
\mathop{}_{\textstyle \longrightarrow}^{\textstyle 31\%} \right] 
\tilde{q}_L \ (688) 
\mathop{}_{\textstyle \longrightarrow}^{\textstyle 32\%}
\neutralinotwo \ (231) 
\mathop{}_{\textstyle \longrightarrow}^{\textstyle 36\%}
\slepton_R \ (157)$.
Left-handed sleptons are heavier than $\neutralinotwo$ and so are
almost never produced.  The $\slepton_R$ events may be isolated with
the cuts of Ref.~\cite{SnowmassLHC}, and the resulting reach is given
by the thick contour of Fig.~\ref{fig:LHCI}.  The reach is, of course,
highly sensitive to the choice of SUSY parameters.  For example, we
also present results for different gluino/squark masses as indicated,
where we have made the naive assumption that the dilepton signal and
background cross sections simply scale with the total squark/gluino
pair production cross section.  We see that for lighter squarks and
gluinos, dramatic improvements result from the increased statistics.
Of course, the results are also dependent on the various branching
ratios, and may be improved by considering both LHC detectors and
multi-year runs.  We see, however, that if sleptons are produced in
gluino and squark cascades, probes of $\widetilde{J}$ to the level of
$10^{-3}$ may be possible.

\subsection{Implications for Models}
\label{sec:implications}

The discovery of slepton flavor or CP violation will have many strong
implications for models.  If slepton flavor violation is discovered,
it will be clear that the stringent low energy constraints on lepton
flavor violation are satisfied not by very massive sfermions or by
sfermion-fermion alignment, but by sfermion degeneracy.  Such a
discovery therefore excludes pure mSUGRA and pure gauge-mediated
models, in which the flavor-blind mediation of SUSY breaking
guarantees the degeneracy of sleptons at a certain mass scale, and
hence renders the slepton mixing matrices trivial.  Such measurements
also have strong implications for models with flavor symmetries, as
they make definite predictions for the $W$ matrices.

At the same time, we will know that the sleptons cannot be completely
degenerate and have mass splittings $\agt \Gamma$.  It is worthwhile
to note that lepton flavor violation is sensitive to mass splittings
at the scale of $\Gamma \approx 0.1 - 1\text{ GeV}$, a level that may
be extremely difficult to achieve by conventional kinematic
techniques.

The discovery of CP violation will have even greater consequences.  As
noted above, CP violation requires not only a significant CP-violating
phase, but also full three generation mixing.  Together with the low
energy constraints, CP violation implies that all three sleptons are
degenerate to $\sim \Gamma$.  These requirements strongly suggest that
the slepton mass matrix has the form
\begin{equation}
m_{\slepton_R}^2 = m_0^2 \left[ {\bf 1} + \left(
\begin{array}{ccc}
\epsilon_{11} & \epsilon_{12} & \epsilon_{13} \\
\epsilon_{21} & \epsilon_{22} & \epsilon_{23} \\
\epsilon_{31} & \epsilon_{32} & \epsilon_{33} \end{array} \right)
\right] \, ,
\end{equation}
where the $\epsilon_{ij}$ are of order $\Gamma/m$ and are of
comparable size for all $i$, $j$.  That any one measurement leads to
such a specific texture is rather remarkable.  Such a texture could
arise in hybrid models, in which the large flavor-blind piece arises
from one source ({\em e.g.}, gauge-mediated SUSY breaking), and the
small $\epsilon_{ij}$ contributions arise from another ({\em e.g.},
supergravity with a large fundamental SUSY-breaking
scale)~\cite{hybrid}.

\section{VERY MASSIVE SUPERPARTNERS}
\label{sec:heavy}

As seen in Sec.~\ref{sec:mixing}, in models beyond the minimal
framework, a wide variety of flavor mixing phenomena may be present,
and, if sfermions are produced at future colliders, the study of such
phenomena may lead to important insights.  It is possible, however,
that some part of the sparticle spectrum will be beyond the reach of
future colliders. In fact, this possibility is realized in a wide
variety of models, and is often found in theories designed to solve
the supersymmetric flavor problem.  These models may be roughly
divided into two categories.  In the first class of models, which we
will refer to as ``heavy QCD models,'' the gluino and all the squarks
are heavy.  Such may be the case in models with gauge-mediated SUSY
breaking, where strongly-interacting sparticles get large
contributions to their masses, in no-scale supergravity, and
generically in models with a heavy gluino, which drives the squark
masses up through renormalization group evolution.  In a second class
of models, ``2--1 models,'' the first and second generation sfermions
have masses ${\cal O}(10 \text{ TeV})$, while the third generation
sfermions are at the weak scale~\cite{2--1}.  Such models are
motivated by the desire to satisfy low energy constraints from, for
example, $K^0 - \bar{K}^0$ mixing and $\mu \to e\gamma$, without the
need for sfermion universality, sfermion alignment, or small
$CP$-violating phases.  At the same time, the extreme fine-tuning
problem arising from very massive third generation sfermions is
alleviated.
 
\subsection{Super-oblique Corrections}
\label{sec:superoblique}

In all of the models described above, the heavy superpartners may well
be beyond the discovery reach of planned future colliders and decouple
from most experimental observables.  However, the mass scale, and
possibly even other properties, of such a sector may be probed by a
class of precision measurements we now
discuss~\cite{CFP1,CFP2,RKS,NPY,PT}.  These probes rely on the fact
that SUSY alone, without further assumptions, already provides
stringent constraints on the form of the low energy theory.  In
particular, let us denote the standard model gauge couplings by $g_i$
and let $h_i$ be their supersymmetric analogues, the
gaugino-fermion-sfermion couplings, where the subscript $i=1,2,3$
refers to the U(1), SU(2), and SU(3) gauge groups.  SUSY implies that
the relations
\begin{equation}
\label{equivalence}
g_i=h_i
\end{equation}
hold to all orders in the limit of unbroken SUSY.  However,
SUSY-breaking mass differences within supermultiplets with standard
model quantum numbers lead to corrections to Eq.~(\ref{equivalence})
that grow logarithmically with the superpartner masses.  Such
deviations from Eq.~(\ref{equivalence}) are thus unambiguous,
non-decoupling, model-independent signals of SUSY-breaking mass
splittings, and by precisely measuring such deviations in processes
involving accessible superparticles, bounds on the mass scale of the
kinematically inaccessible sparticles may be determined.

The corrections to Eq.~(\ref{equivalence}) are highly analogous to the
oblique corrections of the standard model~\cite{CFP1,RKS}.  In the
standard model, non-degenerate SU(2) multiplets lead to inequivalent
renormalizations of the propagators of the $(W,Z)$ vector multiplet,
inducing non-decoupling effects that grow with the mass splitting.
Similarly, in supersymmetric models, non-degenerate supermultiplets
lead to inequivalent renormalizations of the propagators within each
(gauge boson, gaugino) vector supermultiplet, inducing non-decoupling
effects that grow with the mass splitting.  We will therefore refer to
the latter effects as ``super-oblique corrections'' and parametrize
them by ``super-oblique parameters''~\cite{CFP1}.  These corrections
are particularly important because they receive additive contributions
from every split supermultiplet and so may be significantly enhanced.
Furthermore, the simple nature of the corrections allows one to bound
them with many processes in a model-independent fashion.

The coupling constant splittings between $g_i$ and $h_i$ result from
differences in wavefunction renormalizations, and so are most
analogous to the oblique parameter $U$.  We therefore define
\begin{equation}
\susyU_i \equiv h_i / g_i - 1 \ ,
\end{equation}
where the subscript $i$ denotes the gauge group.  These parameters
receive corrections from superpartners of the standard model
particles, and also from possible exotic supermultiplets. For example,
we find~\cite{CFP1}
\begin{eqnarray}
\susyU_1 &\approx& 0.35\% \ (0.29\%) \times \ln R \\
\susyU_2 &\approx& 0.71\% \ (0.80\%) \times \ln R \\
\susyU_3 &\approx& 2.5\%  \times \ln R 
\end{eqnarray}
for 2--1 models (heavy QCD models), where $R = M/m$ is the ratio of
heavy to light mass scales.\footnote{Finite corrections may be
absorbed through a small shift in this definition of $R$.} (In heavy
QCD models, the gluino is decoupled, and so no measurement of
$\susyU_3$ is possible.) We see that measurements of these parameters
at the percent level may be able to detect variations from exact SUSY,
and may even be able to measure the heavy scale $M$. Contributions
from vector-like (messenger) sectors have also been
calculated~\cite{CFP1}, and were found to be significant only for
highly split supermultiplets with masses $\alt {\cal O}(100\text{
TeV})$.

\subsection{Measurements of Super-oblique Parameters}
\label{sec:measurements}

Super-oblique corrections are present in all processes involving
gauginos, and so may be measured at colliders in a variety of ways,
depending on what sparticles are available for study.  The possible
observables at both future lepton and hadron colliders were
systematically classified in Ref.~\cite{CFP2}, where detailed and
representative studies of each of the three super-oblique parameters
were also presented.

\begin{figure}
\centerline{\psfig{file=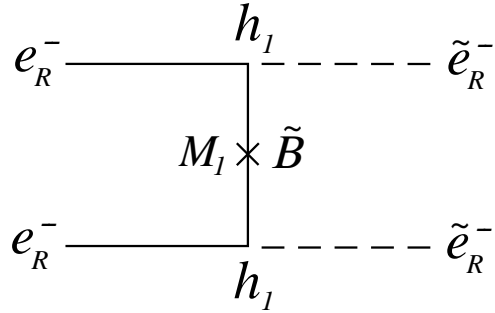,width=0.48\textwidth}}
\vspace*{-.3in}
\caption{Selectron pair production at an $e^-e^-$ collider.
\label{fig:e-e-}}
\vspace*{-.2in}
\end{figure}

Here we will focus on a test of the U(1) couplings, using selectron
production at a future linear collider. We will consider the $e^-e^-$
option at such machines, where a number of beautiful properties make
an extremely precise measurement possible.  Let us consider first
right-handed selectrons.  At an $e^-e^-$ collider, $\selectron_R$
production takes place only through $t$-channel neutralino exchange
(see Fig.~\ref{fig:e-e-}).
Note that while the SUSY process we are interested in is allowed
because the neutralino is a Majorana fermion, many other would-be
backgrounds are absent; for example, $W^- W^-$ pair production is
forbidden by total lepton number conservation.  The backgrounds that
do remain, such as $e^- \nu W^-$, are small and may also be further
suppressed by polarizing both beams.

\begin{figure}[t]
\centerline{\psfig{file=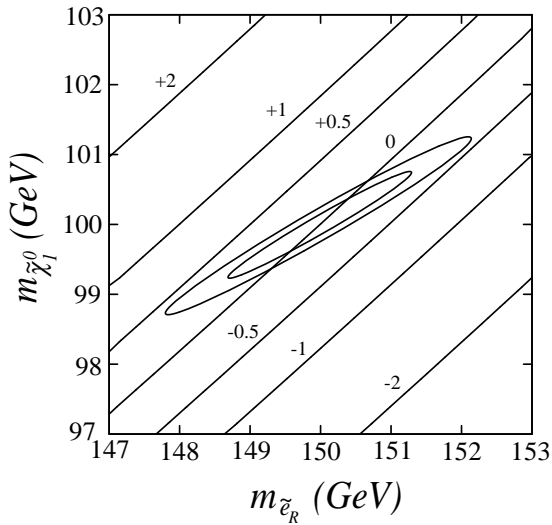,width=0.48\textwidth}}
\vspace*{-.3in}
\caption{The allowed regions, ``uncertainty ellipses,'' of the 
($m_{\selectron_R}$, $m_{\LSP}$) plane, determined by measurements of
the end points of final state electron energy distributions with
uncertainties $\Delta E= 0.3$ GeV and 0.5 GeV.  The underlying central
values are $(m_{\selectron_R}, m_{\LSP}) = (150 \text{ GeV}, 100
\text{ GeV})$, and $\protect\sqrt{s}=500$ GeV.  We also superimpose 
contours (in percent) of the fractional variation of $\sigma_R$ with
respect to its value at the underlying parameters. 
\label{fig:ellipse}}
\vspace*{-.25in}
\end{figure}

The cross section possesses a number of important properties.  First,
it tends to be large relative to the $e^+e^-$ cross section. For
$m_{\selectron_R} \approx 150\text{ GeV}$, $M_1 = 100\text{ GeV}$, and
a total integrated luminosity of $50 \text{fb}^{-1}$, the statistical
error is only 0.3\%.  Second, the cross section is proportional to
$h_1^4$ and is therefore highly sensitive to super-oblique
corrections.  Third, and most importantly, a Majorana mass insertion
in the neutralino propagator is needed.  This greatly reduces
theoretical systematic errors, which are typically dominated by
uncertainties in the $\selectron_R$ and $\LSP$ masses. These masses
are constrained by electron energy distribution endpoints.  The
resulting allowed masses are positively correlated; for example,
typical allowed regions for a year's worth of data and the SUSY
parameters above are given by the ellipses in Fig.~\ref{fig:ellipse}.
Contours of constant cross section are also given.  We see that, since
the cross section decreases for increasing $m_{\selectron_R}$, but
increases for increasing $M_1$ because of the Majorana mass insertion,
the contours of constant cross section run nearly parallel to the
major axes of the ellipses.  Thus, the systematic error on the cross
section from the mass measurements is less than 0.5\%.

Combining these errors, we find that the cross section can be measured
to $\alt 0.6\%$, and so the super-oblique parameter $\susyU_1$ can be
measured to $\alt 0.15\%$.  Such a measurement constrains the heavy
superpartner scale $M$ to within a factor of $1.5$.  For left-handed
selectrons, the dominant diagram will be $\widetilde{W}$ exchange, and
so the cross section will be proportional to $h_2^4$.  The heavy
superpartner scale may then be constrained even more accurately.

Of course, at the percent level, experimental systematic errors may
become important; such uncertainties depend on collider designs and
are subjects of current investigation.  It is clear, however, that if
selectrons are produced at future colliders, the prospects for
precision measurements of super-oblique parameters are indeed
promising.  Measurements of $\susyU_2$ from chargino
production~\cite{FMPT,CFP2} and selectron production~\cite{NPY} in the
more conventional $e^+e^-$ mode of linear colliders, and of $\susyU_3$
from squark branching ratios~\cite{CFP2} have also been studied, with
generally weaker but still promising results.

\subsection{Implications for Models}
\label{sec:comments}

The implications of measurements of the super-oblique parameters
depend strongly on what scenario is realized in nature.  As we have
seen, if some number of superpartners are not yet discovered, bounds
on the super-oblique parameters may lead to bounds on their mass
scale.  In addition, if measurements of more than one super-oblique
parameter may be made, some understanding of the relative splittings
in the heavy sector may be gained.  Inconsistencies among the measured
values of the different super-oblique parameters could also point to
additional inaccessible exotic particles with highly split multiplets
that are not in complete representations of a grand-unified group.
All such insights will provide important input for model building.

If, on the other hand, all superpartners of the standard model
particles are found, the consistency of all super-oblique parameters
with zero will be an important check of the supersymmetric model with
minimal field content.  If instead deviations of the super-oblique
parameters from zero are found, such measurements will provide
exciting evidence for exotic (messenger) sectors with highly split
multiplets not far from the weak scale~\cite{CFP1}.  These insights
could also provide a target for future sparticle searches, and could
play an important role in evaluating future proposals for colliders
with even higher energies, such as the muon collider or higher energy
hadron machines.

\section{CONCLUSIONS}
\label{sec:conclusions}

The discovery of supersymmetry will be just the beginning of a long
and exciting road toward determining which supersymmetric theory is
realized in nature.  Our understanding of the variety of possible
supersymmetric theories has grown dramatically in recent years, and it
is important to determine what colliders and what techniques may
provide stringent tests to distinguish such models. In this talk, I
have discussed examples of phenomena that are generic in supersymmetry
beyond the minimal supergravity framework and may in fact be very
useful for differentiating the vast array of possible models.  Future
colliders may be able to provide stringent probes of sfermion flavor
and CP violation, well beyond current low energy bounds.  In addition,
as many well-motivated models predict that some superpartners are
beyond the discovery reach of future colliders, we have described
methods of probing these sectors through non-decoupling
``super-oblique corrections.''  Just as the oblique corrections of the
standard model provide strong constraints on technicolor models and
other extensions of the standard model, the super-oblique parameters
may provide powerful constraints otherwise inaccessible physics, and
may also have wide implications for theories beyond the minimal
supersymmetric standard model.

\vspace*{-0.03in}

\section*{ACKNOWLEDGEMENTS}

I thank N.~Arkani-Hamed, H.-C.~Cheng, L.~Hall, and N.~Polonsky for
collaborations upon which this talk was based.  This work was
supported in part by the Director, Office of Energy Research, Office
of High Energy and Nuclear Physics, Division of High Energy Physics of
the U.~S.~Department of Energy under Contract DE--AC03--76SF00098 and
in part by the NSF under grant PHY--95--14797.

\end{document}